\documentstyle[12pt,epsf,epsfig,fleqn]{article}
\newcommand{\fips}[1]{\epsffile{#1.eps}}
\newcommand{\be}{\begin{equation}}
\newcommand{\ee}{\end{equation}}
\newcommand{\lee}[1]{\label{#1} \end{equation}}
\newcommand{\bea}{\begin{eqnarray}}
\newcommand{\leea}[1]{\label{#1} \end{eqnarray}}
\newcommand{\eea}{\end{eqnarray}}
\newcommand{\nn}{\nonumber}
\newcommand{\da}{^{\dagger}}
\newcommand{\eq}[1]{eq.~(\ref{#1})}

\newcommand{\eqs}[2]{eqs.~(\ref{#1}) and (\ref{#2})}

\newcommand{\loadeps}[1]{\epsfig{file=#1.eps,width=45mm}}
\newcommand{\floadeps}[1]{\epsfig{file=#1.eps,width=20mm}}
\newcommand{\al}{\alpha}

\newcommand{\ga}{\gamma}
\newcommand{\de}{\delta}

\newcommand{\et}{\eta}

\newcommand{\la}{\lambda}

\newcommand{\De}{\Delta}
\newcommand{\La}{\Lambda}

\begin{document}
\title{Flow equations for QED in the light front dynamics
\vspace*{.5cm}}
\author{E.~L.~Gubankova\thanks{E-mail address: 
elena@hal5000.tphys.uni-heidelberg.de},
 F.~Wegner \vspace*{.5cm}\\
\normalsize\it Institut f\"ur Theoretische Physik der Universit\"at 
Heidelberg \\
\normalsize\it Philosophenweg 19, D69120 Heidelberg, FRG}

\maketitle

\vspace*{1cm}
\begin{abstract}

The method of flow equations is applied to QED on  the light front.
Requiring that the partical number conserving terms in the Hamiltonian
are considered to be diagonal and the other terms off-diagonal
an effective Hamiltonian is obtained which reduces the positronium problem
to a two-particle problem, since the particle number violating contributions
are eliminated. No infrared divergences appear. The ultraviolet 
renormalization can be performed simultaneously.

\end{abstract}

\newpage
\section{Introduction}

We use the method of flow equations to construct an effective
Hamiltonian starting from the light-front formulation which can be
used to solve the bound state problem.

Recently Glazek and Wilson \cite{GlWi} and one of the authors \cite{We} 
have suggested methods to diagonalize Hamiltonians continuously which have
been called similarity renormalization and flow equations by the authors, 
resp.

It is common to both methods that they eliminate by means of a unitary
transformation initially the off-diagonal matrix elements between
states with large energy differences and continue with states closer and 
closer 
in energy, so that off-diagonal matrix elements between states of
energy difference larger than $\la$ are eliminated or strongly suppressed. 
The final aim is to eliminate them completely ($\la \rightarrow 0$)
and to obtain a diagonalized Hamiltonian.

Application of this method to an $n$-orbital model has shown, that a literal
use of this concept can lead to convergency problems \cite{We}. Instead one
may leave the idea of diagonalizing immediately in favor of 
block-diagonalizing.
If matrix-elements between states of equal particle number are considered
diagonal, then the procedure brings the Hamiltonian into a block-diagonal 
form, a procedure which worked much better in the case of the $n$-orbital
model, where block-diagonalization with respect to the quasiparticle
number (number of electrons above the Fermi edge plus number of holes
below the Fermi edge) was performed.

It becomes apparent from the calculations by Jones, Perry and Glazek 
\cite{JoPeGl} on the basis of the similarity transformation,
that this scheme works 
well down to energy differences of the order of Rydberg, but if one
goes below, then contributions in higher orders in the coupling
become important.

Indeed it seems to be rather difficult to obtain bound states from
plane waves by continuous unitary transformations. In eliminating only
the terms which do not conserve the number of particles one postpones
the diagonalization, but reduces the problem to one in the space
of fixed particle number \cite{HaOk}. Thus for the positronium problem it is 
sufficient
to determine the one- and two-particle contribution of the Hamiltonian
for electrons and positrons.

Basically the procedure is very similar to that of the elimination
of the electron-phonon interaction \cite{LeWe} which yields an effective
attractive interaction between electrons responsible for superconductivity.
Both the method of flow equations and the similarity remormalization \cite{Mi}
yield results different from Fr\"ohlich's original ones \cite{Fr} but in very
good agreement with more sophisticated methods. In QED it is the interaction
of the electrons with the photons instead of the phonons which has to be 
eliminated.

A basic advantage of the methods of similarity renormalization and flow 
equations in comparison to conventional perturbation theory is, that one
obtains normally less singular effective interactions.

This procedure is similar to the Tamm-Duncoff Fock space truncation 
\cite{TaDa,Pa1,Pa2,TrPa} 
in the sense that also in this truncation particle number changing interactions
are eliminated.

We define an effective, renormalized Hamiltonian 
\be
H_{eff}=\lim_{\la\rightarrow 0,\La\rightarrow\infty} H_B(\la,\La)
\lee{in1}
where $H_B(\la,\La)$ is obtained by means of the unitary transformation
$U(\la,\La)$ from the bare Hamiltonian (regularized at the cutoff $\La$)
\be
H_B(\la,\La)=U(\la,\La)H_B(\La)U\da (\la,\La).
\lee{in2}
The unitary transformation $U(\la,\La)$ is determined by 
the flow equations below.

Flow equations eliminate the far-off diagonal matrix elements 
in the 'energy space' \footnote{Under the 'energy space' we understand
the basis of the free Hamiltonian $H^{(0)}_d$, corresponding to the sum
of kinetic energies of single noninteracting particles.}
(the matrix elements between the states with large energy jumps $|E_i-E_j|>
\la$,
where $\la$ is the 'running' UV cutoff) but only for those blocks that change
the number of quasiparticles. The value of
$\la=\La\rightarrow\infty$ corresponds to the initial bare Hamiltonian,
a finite $\la$ determines the effective Hamiltonian at an intermediate stage,
for $\la=0$ the elimination of the 'non-diagonal' part in the Fock
representation is complete fig. (1).

The method of flow equations in the perturbative theory frame is discussed
in section 2. In section 3 we apply this scheme to QED on the light front.
The effective Hamiltonian serves for the calculation of bound states,
which due to the triviality of the vacuum state is now much simpler.

\section{Flow equations in the perturbative frame}

The flow equation reads
\be
\frac{dH(l)}{dl}=[\eta(l),H(l)].
\ee
This is the differential form of a continuous unitary transformation
acting on the Hamiltonian in general; $\eta(l)$ is antihermitean generator
of the transformation and $l$ is the flow parameter.
The aim of the flow equations is to bring the initial
Hamiltonian matrix $H(l=0)$
to a diagonal (or block-diagonal) form. Finite values of the flow parameter
$l$ correspond to intermediate stages of diagonalization
with the band-diagonal structure of Hamiltonian (similarity renormalization) 
\cite{GuWe}.
The result of the procedure at $l\rightarrow\infty$ is a
block-diagonal Hamiltonian.
Also the transformation is designed in a way to avoid
small energy denominators usually present in perturbation theory.

What is the choice of the generator $\eta(l)$ that performs such a
transformation? Break the Hamiltonian into a 'diagonal' and a 'rest'
part $H=H_d+H_r$. Then the prescription in \cite{We} is
\be
\eta(l)=[H_d(l),H_r(l)].
\ee
At this step the unitary transformation is defined. The only 
freedom left is the principle of separation into the 'diagonal' and 
the 'rest' part. It depends on the problem one wants to treat.

Our goal is to transform the Hamiltonian into blocks with the 
same number of (quasi)-particles. 
This means, that we define the 'diagonal' part $H_d$ as the part 
of the interaction which conserves the number of particles
(electrons, positrons, photons),
and the 'rest' $H_r$ as the particle number changing part.
In the case of QED(QCD),
where the electron-photon (quark-gluon) coupling is present,
the number of photons (gluons) is conserved in each block of the 
final effective Hamiltonian.

As a result of the unitary transformation new interactions are induced
(see below). They are absent at $l=0$ and are generated as $l$ increases.
They also give rise to new terms 
in the generator of transformation $\eta(l)$. This in its turn
generates new interactions again.

To be able to perform the calculations analytically
we proceed in a perturbative frame and truncate
the series assuming the coupling constant is small.

As illustration of the method we consider QED on the light front.
For any finite value of $l$ one has
\be
H(l)=H_d^{(0)}+H_r^{(1)}+H_d^{(2)}+H_r^{(2)}+...
\ee
where the superscript denotes the order in the bare coupling constant,
$H^{(n)}\sim e^n$; the indices 'd' and 'r' indicate the diagonal and the
rest parts correspondingly.
The part $H_d^{(0)}$ is the free Hamiltonian, corresponding to the single
particle energies with the structure in secondary quantization
$a\da a,b\da b, d\da d$, where $a,b,d$ are the annihilation operators
of the photons, electrons and positrons correspondingly; 
$H_r^{(1)}$ denotes the electron-photon coupling (of the type $a\da b\da b$);
$H_d^{(2)}$ is the second order diagonal part of the Hamiltonian,
having the structure $b\da d\da bd, b\da b\da bb, d\da d\da dd$
(in the light front
they correspond to the canonical instantaneous (seagull) and to newly 
generated interactions in the diagonal sector in second order 
\cite{GuWe2},\cite{ZhHa}).
Note, that the diagonal part in the flow equations is not only the free
Hamiltonian but the full particle number conserving part of the 
effective Hamiltonian.
The choice of only $H_d^{(0)}$ as the diagonal part gives rise
to the band-diagonal structure of the effective Hamiltonian
in each 'particle number' sector in the similarity renormalization 
scheme \cite{GuWe}. However, this makes a difference for the diagonal
part only if one goes beyond third order in $e$.

The generator of the transformation is
\be
\eta(l)=[H_d,H_r]=[H_d^{(0)},H_r^{(1)}]+[H_d^{(0)},H_r^{(2)}]+...=
\eta^{(1)}+\eta^{(2)}+...
\ee
Up to second order the flow equation reads
\be
\frac{dH(l)}{dl}=[\eta,H]=
[[H_d^{(0)},H_r^{(1)}],H_d^{(0)}]+[[H_d^{(0)},H_r^{(1)}],H_r^{(1)}]+
[[H_d^{(0)},H_r^{(2)}],H_d^{(0)}]+...
\ee
Also terms of higher orders in $e$ are generated by the flow equations.

In the basis of the eigenfunctions of the free Hamiltonian
$H_d^{(0)}$
\be
H_d^{(0)}|i>=E_i|i>
\ee
one obtains for the matrix-elements between the many-particle states
\bea
&& \eta_{ij}=(E_i-E_j)H_{rij}^{(1)}+(E_i-E_j)H_{rij}^{(2)}+...\nn\\
&& \frac{dH_{ij}}{dl}=-(E_i-E_j)^2H_{rij}^{(1)}
+[\eta^{(1)},H_r^{(1)}]_{ij}-(E_i-E_j)^2H_{rij}^{(2)}+...
\leea{fe7}
The energy differences are given by
\bea
&& E_i-E_j = \sum_{k=1}^{n_2}E_{i,k}-\sum_{k=1}^{n_1}E_{j,k}
\leea{fe9}
where $E_{i,k}$ and $E_{j,k}$ are the energies of the created
and annihilated particles, respectively.

The energy $E_i$ depends on the flow parameter $l$ only in second
order in the coupling. Therefore one has
\bea
&& \frac{dH_{rij}^{(1)}}{dl}=-(E_i-E_j)^2H_{rij}^{(1)}\nn\\
&& H_{rij}^{(1)}(l)=H_{rij}^{(1)}(l=0){\rm e}^{-(E_i-E_j)^2l}
=H_{rij}^{(1)}(\la=\La\rightarrow\infty){\rm e}^{-\frac{(E_i-E_j)^2}{\la^2}}
\eea
Here we have used the physical meaning of the flow parameter $l$.
Namely, in the similarity renormalization scheme it defines the width of 
the band $\la$,
corresponding to the UV-cutoff, where the matrix elements of
the effective Hamiltonian are not zero ($|E_i-E_j|<\la$) \cite{GuWe}.
The connection between these two quantities is
\be
l=\frac{1}{\la^2}
\ee
In the flow equations as used here the matrix elements of the interactions, 
which change the number
of particles, are strongly suppressed, if the energy difference exceeds $\la$,
 while for the particle number
conserving part of the effective Hamiltonian the matrix elements with
all energy differences are present.

In the flow equations $\la$ characterizes the smooth UV-cutoff.
This fact insures the analytical behavior of the effective
Hamiltonian with $\la$, that helps in numerical calculations.

In second order one has to distinguish between the behavior of the 'diagonal'  
and the 'rest' term. For the 'rest' part one has
\be
\frac{dH_{rij}^{(2)}}{dl}=[\eta^{(1)},H^{(1)}]_{rij}
-(E_i-E_j)^2H_{rij}^{(2)},
\ee
where index 'r' by $[\eta^{(1)},H^{(1)}]_{r}$ defines
the particle number changing part of the commutator. Introduce
\be
H_{rij}^{(2)}(l)={\rm e}^{-(E_i-E_j)^2l}\tilde{H}_{rij}^{(2)}(l).
\ee
Then the solution reads
\be
\tilde{H}_{rij}^{(2)}(l)=\tilde{H}_{rij}^{(2)}(l=0)
+\int_0^l dl'{\rm e}^{(E_i-E_j)^2l'}[\eta^{(1)},H^{(1)}]_{rij}(l').
\ee
For the 'diagonal' part one has
\be
\frac{dH_{dij}^{(2)}}{dl}=[\eta^{(1)},H^{(1)}]_{dij}
\ee
and the solution is
\be
H_{dij}^{(2)}(l)=H_{dij}^{(2)}(l=0)
+\int_0^l dl'[\eta^{(1)},H^{(1)}]_{dij}(l').
\ee
Note, that though in general the commutator $[[H_d^{(0)},H_d^{(2)}],H_d^{(0)}]$
is not zero, it is not present in the flow equation due to the definition 
of the diagonal part. The corresponding commutator 
$[[H_d^{(0)},H_r^{(2)}]H_d^{(0)}]$ in the 'non-diagonal' sector insures the 
band-diagonal
form for the 'rest' interaction and also gives rise to the different
structure of the generated interaction (the integral term) in the 'rest' 
and 'diagonal' sectors. 

The commutator $[\eta^{(1)},H^{(1)}]$ gives rise to new terms 
in second order in the bare coupling $e$. In the case of QED it induces
new types of interactions and generates the renormalization group corrections
to the electron (photon) masses. The coupling constant starts to run 
in third order in $e$.

Note, that the method of flow equations (and also the similarity 
renormalization) enables one to build 
an effective low energy Hamiltonian together with all, 'canonical' and 
'new' \cite{Pe} counterterms, 
found (from the coupling coherence condition 
\cite{Pe},\cite{PeWi})
order by order in the coupling constant $e$. This defines an effective
renormalized Hamiltonian which may be used for the numerical solution 
of the bound state problem.

\section{Renormalized effective electron-positron interaction}

In this section we give the effective Hamiltonian in the light front
dynamics for the positronium system, generated by the unitary transformation
in section 2, compare \cite{GuWe2}.

The light front Schr\"odinger equation for the positronium model reads
\be
H_{LC}|\psi_n>=M_n^2|\psi_n>
\lee{re1}
where $H_{LC}=P^{\mu}P_{\mu}$ is the invariant mass (squared) operator, 
refered for convenience to as the light front Hamiltonian of positronium 
and $|\psi_n>$ being the corresponding eigenfunction; $n$ labels all
the quantum numbers of the state. 

The canonical Hamiltonian of the system
$H_{LC}$ contains infinitely many Fock sectors (i.e. one has for 
the positronium  wave function $|\psi_n>=c_{e\bar{e}}|(e\bar{e})_n>
+c_{e\bar{e}\ga}|(e\bar{e}\ga)_n>
+c_{e\bar{e}\ga\ga}|(e\bar{e}\ga\ga)_n>+...$) and each Fock sector contains
states with arbitrarily large energies. We now

$(1)$ introduce the bare cutoff (regularization) with the result
$H_{LC}^B(\La)$ - the bare Hamiltonian;

$(2)$ perform the unitary transformation by means of flow equations
\eqs{in1}{in2} with the result $H_{LC}^{eff}$ - the effective renormalized
Hamiltonian (table 1 for finite value of $\la$);

$(3)$ truncate the Fock space to the lowest Fock sector ($|e\bar{e}>$)
with the result $\tilde{H}_{LC}^{eff}$ - the effective 
renormalized Hamiltonian acting in the electron-positron sector.

Then the eigenvalue equation reads
\be
\tilde{H}_{LC}^{eff}|(e\bar{e})_n>=M_n^2|(e\bar{e})_n>.
\lee{re1a}
The effective light front Hamiltonian consists of
the free (noninteracting) part and the effective electron-positron interaction
\be
\tilde{H}_{LC}^{eff}=H_{LC}^{(0)}+V_{LC}^{eff}
\lee{re2}
The light front equation \eq{re1a} 
is then expressed by the integral equation (the coordinates are given in
fig. (2)
\bea
&& \hspace{-2cm} \left(\frac{m^2+\vec{k}^{'2}}{x'(1-x')}-M_n^2\right)
\psi_n(x',\vec{k}_{\perp}^{'};s_3,s_4)\nn\\
&+&\sum_{s_1,s_2}\int_D \frac{dx d^2 k_{\perp}}{2(2\pi)^3}
<x',\vec{k}_{\perp}^{'};s_3,s_4|V_{LC}^{eff}|x,\vec{k}_{\perp};s_1,s_2>
\psi_n(x,\vec{k}_{\perp};s_1,s_2)=0
\leea{re3}
The integration domain $D$ is restricted by the covariant cutoff
condition of Brodsky and Lepage \cite{BrLe}
\be
\frac{m^2+\vec{k}^{2}}{x(1-x)}\leq \La^2+4m^2
\lee{re3a}
which allows for states which have a kinetic energy below the 
cutoff $\La$.
 
For the effective electron-positron interaction one has 
in the exchange and annihilation channels 
\be
V_{LC}^{eff}=V_{exch}+V_{ann}=
\sum_{channel}\lim_{\la\rightarrow 0 }
(V_{\la}^{gen}+V_{\la}^{inst}+V_{\la}^{PT})
\lee{re4}
where the terms in \eq{re4} correspond to generated, instantaneous and 
perturbative photon exchange interactions in the electron-positron sector. 
The interactions generated by the flow equations 
from the Hamiltonian in the light front frame \cite{GuWe2} are,
fig. (2)

in the {\bf exchange channel}
\bea
V_{\la}^{gen} &=& -e^2N_1
\left(\frac{\int_{\la}^{\infty}\frac{df_{\la'}(\De_1)}{d\la'}f_{\la'}
(\De_2)d\la'}{\tilde{\De}_1}+
\frac{\int_{\la}^{\infty}\frac{df_{\la'}(\De_2)}{d\la'}f_{\la'}(\De_1)d\la'}
{\tilde{\De}_2}\right)\nn\\
V_{\la}^{inst} &=& -\frac{4e^2}{(x-x')^2}\de_{s_1s_3}\de_{s_2s_4}\nn\\
V_{\la}^{PT} &=& -e^2N_1
\frac{1}{\tilde{\De}_3}f_{\la}(\De_1)f_{\la}(\De_2)
\leea{re5}

in the {\bf annihilation channel} 
\bea
V_{\la}^{gen} &=& e^2N_2
\left(\frac{\int_{\la}^{\infty}\frac{df_{\la'}(M_0^2)}{d\la'}f_{\la'}
(M_0^{'2})d\la'}{M_0^2}+
\frac{\int_{\la}^{\infty}\frac{df_{\la'}(M_0^{'2})}{d\la'}f_{\la'}(M_0^2)d\la'}
{M_0^{'2}}\right)\nn\\
V_{\la}^{inst} &=& 4e^2\de_{s_1\bar{s}_3}\de_{s_2\bar{s}_4}\nn\\
V_{\la}^{PT} &=& e^2N_2
\frac{1}{M_n^2}f_{\la}(M_0^2)f_{\la}(M_0^{'2}).
\leea{re6}
where $f_{\la}(\De)$ is the similarity function
\be
f_{\la}(\De) = e^{-\frac{\De^2}{\la^4}}
\ee
where we have used the rescaled value of the UV cutoff 
$\la \rightarrow \frac{\la^2}{P^+}$.

As long as $\la$ is finite we have taken the still existing terms in
$H_r^{(1)}$ into account by the perturbative result obtained by Lepage
and Brodsky \cite{BrLe}.

The functions $N_1, N_2$ (current-current terms) and 
the energy denominators $\tilde{\De}_i, i=1,2,3$, $M_0^2,M_0^{'2},M_n^2$ 
are defined in the light front dynamics \cite{ZhHa} 
as follows \cite{GuWe2}  fig. (2)
\bea
N_1&=&\delta_{s_1s_3}\delta_{s_2s_4}
 T_1^{\bot}\cdot T_2^{\bot}
 -\delta_{s_1\bar{s}_2}\delta_{s_1\bar{s}_3}\delta_{s_2\bar{s}_4}
 2m^2\frac{(x-x')^2}{xx'(1-x)(1-x')}\nonumber\\
&&+im\sqrt{2}(x'-x) \left[ \delta_{s_1\bar{s}_3}\delta_{s_2s_4}
 \frac{s_1}{xx'}T_1^{\bot}\cdot \varepsilon_{s_1}^{\bot}
 +\delta_{s_1s_3}\delta_{s_2\bar{s}_4}
 \frac{s_2}{(1-x)(1-x')}T_2^{\bot}\cdot \varepsilon_{s_2}^{\bot} \right]
\nn\\
N_2&=&\delta_{s_1\bar{s}_2}\delta_{s_3\bar{s}_4}
 T_3^{\bot}\cdot T_4^{\bot}
 +\delta_{s_1s_2}\delta_{s_3s_4}\delta_{s_1s_3}
 2m^2\frac{1}{xx'(1-x)(1-x')}\nonumber\\
&&+im\sqrt{2} \left[ \delta_{s_3\bar{s}_4}\delta_{s_1s_2}
 \frac{s_1}{x(1-x)}T_3^{\bot}\cdot \varepsilon_{s_1}^{\bot}
 -\delta_{s_3s_4}\delta_{s_1\bar{s}_2}
 \frac{s_3}{x'(1-x')}T_4^{\bot}\cdot \varepsilon_{s_4}^{\bot *} \right] \nn \\
&& \varepsilon_s^i = -\frac{1}{\sqrt{2}}(s, i)
\leea{re7}
and
\bea
T_1^i&=&- \left[ 2\frac{(k_{\bot}-k'_{\bot})^i}{(x-x')}+\frac{k_{\bot}^i(s_2)}{
(1-x)}+
 \frac{k_{\bot}^{'i}(\bar{s}_2)}{(1-x')} \right] \; ; \qquad
T_2^i=2\frac{(k_{\bot}-k'_{\bot})^i}{(x-x')}-\frac{k_{\bot}^i(s_1)}{x}-
 \frac{k_{\bot}^{'i}(\bar{s}_1)}{x'} \nonumber \\
T_3^i&=&-\frac{k_{\bot}^{'i}(\bar{s}_3)}{x'}
 +\frac{k_{\bot}^{'i}(s_3)}{(1-x')} \; ; \qquad
T_4^i=\frac{k_{\bot}^i(\bar{s}_1)}{(1-x)}
 -\frac{k_{\bot}^i(s_1)}{x}\nonumber\\
&& k_{\bot}^i(s) = k_{\bot}^i+is\varepsilon_{ij}k_{\bot}^j \; ; \qquad
 \varepsilon_{ij}=\varepsilon_{ij3} \; ; \qquad
 \bar{s} = -s \nn 
\leea{re8}
with the definitions
\bea
\tilde{\De}_1 &=& \frac{(xk'_{\bot}-x'k_{\bot})^2+m^2(x-x')^2}{xx'}\; ; \qquad
 \tilde{\De}_2=\tilde{\De}_1|_{x\rightarrow(1-x),x'\rightarrow(1-x')} \nn \\
&&\De_1=\frac{\tilde{\De}_1}{x'-x} \; ; \qquad
 \De_2=\frac{\tilde{\De}_2}{x'-x} \nn \\
\tilde{\De}_3 &=& \frac12(\tilde{\De}_1+\tilde{\De}_2)+|x-x'|
 \left( \frac{1}{2}(M_0^2+M_0^{'2})-M_n^2 \right) \nn \\
&&M_0^2=\frac{k_{\bot}^2+m^2}{x(1-x)} \; ; \qquad
 M_0^{'2}=\frac{k_{\bot}^{'2}+m^2}{x'(1-x')} \nn \\
&&P^-=\frac{(P^{\bot})^2+M_n^2}{P^+} \; ; \qquad P=(P^+,P^{\bot}) \; ; \qquad
 M_n = 2m + B_n
\; . \leea{re9}
Here $x$ is the light front fraction of the electron momentum,
$P$ is the total momentum of positronium and $B_N$ the binding energy of the
positronium.

Explicitly the integrals in \eqs{re5}{re6} are given
\be
 \int_{\la}^{\infty}\frac{df_{\la'}(\De_1)}{d\la'}f_{\la'}(\De_2)d\la'
= \frac{\tilde{\De}_1^2}{\tilde{\De}_1^2+\tilde{\De}_2^2}
(1-f_{\la}(\De_1)f_{\la}(\De_2)).
\lee{re10}
We use instead of the light front parameterization
fig. (2), the instant form. Namely, we express the variable
$(x,\vec{k}_{\perp})$ in terms of the equal-time variable 
$\vec{k}=(k_z,\vec{k}_{\perp})$ as
\bea
&& x=\frac{1}{2}\left(1 + \frac{k_z}{\sqrt{\vec{k}^2 + m^2}} \right)\\
&& \vec{k}^2=k_z^2+\vec{k}_{\perp}^2
\leea{re11}
and similarly for $x'$ and $\vec{k}^{'2}$ as function of $k_z^{'}$.

Then the electron-positron interaction reads in the exchange
and annihilation channels
\bea
V_{\la} &=& V_{\la,exch}+V_{\la,ann} \nn \\
&=&-e^2N_{1,\la} \left[
\left(\frac{\tilde{\De}_1+\tilde{\De}_2}
{\tilde{\De}_1^2+\tilde{\De}_2^2}\right) 
(1-{\rm e}^{-\frac{\De_1^2+\De_2^2}{\la^4}})
+\frac{1}{\tilde{\De}_3}
{\rm e}^{-\frac{\De_1^2+\De_2^2}{\la^4}} 
\right] \nn \\
&&+\left( -\frac{4 e^2}{(x - x')^2} \: \de_{s_1 s_3} \: 
 \de_{s_2 s_4} \right)\nn\\
&&+e^2N_{2,\la} \left[ 
\left(\frac{M_0^2+M_0^{'2}}{M_0^4+M_0^{'4}}\right)
(1-{\rm e}^{-\frac{M_0^4+M_0^{'4}}{\la^4}})
+\frac{1}{M_n^2}
{\rm e}^{-\frac{M_0^4+M_0^{'4}}{\la^4}}
 \right] \nn \\
&&+\left( 4 e^2
 \: \de_{s_1 \bar{s}_3} \: \de_{s_2 \bar{s}_4} \right)
\leea{re12}
where one has in the instant frame
\be
M_0^2=4(\vec{k}^2+m^2)
\lee{re13}
and similarly for $M_0^{'2}$ as function of $\vec{k}^{'2}$.
For all the quantities, defined in \eqs{re7}{re8}, the substitution 
$x(k_z), x'(k_z^{'})$ is to be done.
To the leading order of the nonrelativistic approximation
$|\vec{k}|/m \ll 1$
one obtains (for the exchange channel)
\bea
&& \tilde{\De}_1\sim\tilde{\De}_2\sim\tilde{\De}_3=\tilde{\De}=
(\vec{k}-\vec{k'})^2\nn\\
&& V_{\la}^{gen}\approx -e^2\frac{N_1}{(\vec{k}-\vec{k'})^2}
(1-f_{\la}^2(\De))\nn\\
&& V_{\la}^{PT}\approx -e^2\frac{N_1}{(\vec{k}-\vec{k'})^2}
f_{\la}^2(\De)\nn\\
&& \De=\frac{(\vec{k}-\vec{k'})^2}{x'-x}
\leea{re14}
This gives for the electron-positron interaction in the whole 
nonrelativistic range of $\la$, $\la\ll m$
\be
V^{|e\bar{e}>}\approx -e^2\frac{N_1}{(\vec{k}-\vec{k'})^2}
- \frac{4e^2}{(x-x')^2}\de_{s_1s_3}\de_{s_2s_4}
\lee{re15}
the $\la$-independent result.
Making use of the following expressions
\bea
&& N_1^{diag}\approx -4\frac{(\vec{k}_{\perp}-\vec{k}_{\perp}^{'})^2}
{(x-x')^2}\de_{s_1s_3}\de_{s_2s_4}\nn\\
&& (\vec{k}-\vec{k'})^2=(\vec{k}_{\perp}-\vec{k}_{\perp}^{'})^2+
(k_z-k_z^{'})^2\approx (\vec{k}_{\perp}-\vec{k}_{\perp}^{'})^2+
4m^2(x-x')^2
\leea{re16}
one obtains in leading order of the nonrelativistic approximation
the $3$-dimensional Coulomb interaction $(e^2=4\pi\al)$
\be
V^{|e\bar{e}>}\approx 16m^2\left( 
 - \frac{e^2}{(\vec{k}-\vec{k'})^2}\right)
\de_{s_1s_3}\de_{s_2s_4}
\lee{re17}
hence the rotational invariance is restored in this order.
This result \eq{re17} is valid for any nonrelativistic value of cutoff $\la$
and does not depend on the details of the similarity function $f_{\la}(\De)$.

We perform the limit of $\la\rightarrow 0$ in the effective interaction, 
\eqs{re5}{re6}, that corresponds
to the complete elimination of the electron-photon vertex. Then the 
perturbative
term for the dynamical photon exchange $V^{PT}$ vanishes.
Therefore the effective interaction \eq{re4}, generated by the flow equations,
is defined in the whole parameter region, (except maybe for 
the Coulomb singularity point $\vec{q}=\vec{k}-\vec{k'}=0$) 
as follows 
\bea
V_{LC}^{eff} &=& V_{exch}+V_{ann} \nn \\
&=&-e^2N_1 
\left(\frac{\tilde{\De}_1+\tilde{\De}_2}
{\tilde{\De}_1^2+\tilde{\De}_2^2}\right) 
+\left( -\frac{4 e^2}{(x - x')^2} \: \de_{s_1 s_3} \: 
 \de_{s_2 s_4} \right)\nn\\
&&+e^2N_2 
\left(\frac{M_0^2+M_0^{'2}}{M_0^4+M_0^{'4}}\right)
+\left( 4 e^2
 \: \de_{s_1 \bar{s}_3} \: \de_{s_2 \bar{s}_4} \right)
\; , \leea{re18}

The effective interaction $V_{LC}^{eff}$ obtained in \eq{re18}
stands for the kernel in the integral equation \eq{re3} for the calculation of
the bound state spectrum and the wave functions of positronium.
The integral equation \eq{re3} with the effective interaction given above
is to be used for the numerical calculations of positronium spectrum. 

This interaction was also used for an analytical analysis of the positronium
ground states.
The standard singlet-triplet mass splitting for positronium 
$\frac{7}{6}\alpha^2 Ryd$ was obtained and the degeneracy of the triplet
ground state $n=1$ was recovered \cite{GuWe2},\cite{JoPeGl}.

Instead of the choice $\et^{(1)}_{ij} = (E_i-E_j)H^{(1)}_{rij}$ one can
choose $\et^{(1)}_{ij} = {\rm sign}(E_i-E_j)H^{(1)}_{rij}$. With this choice
in the above equations the ratio $\frac{\tilde{\De}_1+\tilde{\De}_2}
{\tilde{\De}_1^2+\tilde{\De}_2^2}$ has to be replaced by
$\frac{2}{\tilde{\De}_1+\tilde{\De}_2}$. If one approximates $M_n^2 =
\frac12(M_0^2+M^{'2}_0)$ in the expression for $\tilde{\De}_3$,
then $\tilde{\De}_3$ equals $\frac{\tilde{\De}_1+\tilde{\De}_2}2$, and
one obtains the perturbation theoretic result 
$V^{PT}_{\la =\La \rightarrow \infty}+V^{inst}$ \cite{BrLe} for the
effective interaction. These both interactions with different choices of
$\et$ do not coincide.
Both have a leading Coulomb behavior, but they differ in the order $e^2q^0$,
which determines the fine structure splitting. Quite generally the two particle
interaction is of order $q^{-2}$ and subleading terms in $q$. In the order
of fine structure splitting $\al^4$ also terms of order $e^4q^{-1}$ and 
$e^6q^{-2}$ will in general be important. Thus we expect that the difference 
in order $e^2q^0$ will be compensated by differences in order $e^4$ and $e^6$.
They should yield in total the correct fine structure splitting.

Using the similarity transformation Brisudova and Perry \cite{BrPe} have
obtained the correct
spin-spin interaction for the positronium from the effective interaction in
order $e^2$. This had to be expected, since each spin enters the interaction
with a factor of order $q/m$ as compared to the leading Coulomb
interaction. Thus the two-spin interaction enters only in order $q^0$ or
higher. The only contribution to order $\al^4$ comes from order $e^2$.
The same holds for the spin-triplet splitting (which is quadratic in the
spin), and has been obtained correctly for the ground state by Jones, Perry
and Glazek \cite{JoPeGl}.
By the same reasoning contributions to the spin-orbit coupling are of order
$q^{-1}$, so that contributions from order $e^2$ and $e^4$ have to be expected,
but not from order $e^6$.
Spin independent contributions can be of order $q^{-2}$.
To obtain all contributions of order $\al^4$ one has to consider the
interaction in order $e^2$, $e^4$ and $e^6$.

\section{Conclusions and outlook}

In this work we have outlined a strategy to derive an effective 
renormalized Hamiltonian by means of flow equations. Application of the
flow equations with the condition, that particle number conserving terms
are considered diagonal and those changing the particle number off-diagonal
led as in other cases to a useful effective Hamiltonian.

The main advantage of this procedure as compared 
with the similarity renormalization of Glazek and Wilson \cite{GlWi} 
is, that finally states of
different particle number are completely decoupled, since
the particle number violating contributions are eliminated down to $\la=0$.
Thus one is able to truncate the Fock space and the positronium problem 
reduces to a two particle problem which can be
analyzed further analytically (since in leading order one obtains
the nonrelativistic Coulomb problem) or numerically \cite{Nu}
for positronium bound states.

The effective Hamiltonian, obtained by the similarity transformation,
is band-diagonal in the energy space. The width of the band $\la$
introduces the artificial parameter in the procedure,
which is defined from the physical reasoning
($\la$ is low enough to neglect the contribution of high Fock states,
but is restricted from below to stay in perturbation theory region).
Flow equations as used here with the particle number 
conserving part of Hamiltonian to be diagonal, have no additional
parameter and converge well as $\la\rightarrow 0$ \cite{We}
to the effective Hamiltonian, which is block-diagonal in particle number
and can be used directly for the numerical calculations 
of the spectrum (work in preparation).

The procedure of elimination of nondiagonal blocks, that change
the number of quasiparticles, is performed not just in one step as
in the method of Tamm-Dancoff truncation but rather continuously
for the states with different energies in sequence.
This is the main advantage of the proposed method as compared
with Tamm-Dancoff truncation,
the possibility to perform simultaneously the ultraviolet renormalization
of the initial Hamiltonian.
In general, in the definite order of perturbation theory
all counterterms, associated with canonical operators of the theory
and also with possible new operators induced by unitary transformation,
can be obtained in the procedure \cite{GuWe}. Since different
sectors of the effective Hamiltonian are decoupled, 
one does not encounter the usual difficulties of Tamm-Dancoff truncation
and the methods related to it. Namely, the counterterms to be introduced
are 'sector-' and 'state-' independent (work in preparation and \cite{HaOk}).
 
If one goes beyond
the tree approximation then one obtains terms with ultraviolet divergences
which have to be renormalized. This has not been considered in this
paper. Simultaneously also terms describing interactions between more
than two particles are generated.
In this approach we were not faced with infrared problems
(except infrared collinear divergences along the light front, that are
removed in the considered sector, if all possible diagrams in this order
are taken into account \cite{GuWe2}). 

By means of the flow equation method one can simultaneously 
renormalize the initial field theoretical Hamiltonian
and construct the effective Hamiltonian, for which the Fock space
truncation is valid. In order to solve the flow equations analytically 
we were forced to apply in this work the perturbative theory expansion. 
One is able to improve this approach 
systematically by going to higher orders in the coupling.

We consider flow equations as a method which can also be used beyond
perturbation theory in a self-consistent way.
Examples in solid-state physics are the flow of the tunneling-frequency
in the spin-boson model \cite{KeMiNe} and of the phonon energies
in the electron-phonon coupling \cite{LeWe}. Due to the flow 
the couplings decay even at resonance.

\paragraph{Acknowledgments}
This work was partially supported by the Deutsche Forschungsgemeinschaft,
grant no. GRK 216/5-95.

\newpage

Figure 1: Flow equations perform the block-diagonalization of the bare 
Hamiltonian of the canonical theory $H_B(\La)$ into a Hamiltonian consisting
of blocks with equal number of particles. For a finite value of $\la$ the
matrix elements of the 'particle number changing' sectors are squeezed into an 
energy band with roughly $|E_i-E_j|<\la$ (left hand side picture) and are 
eliminated completely as $\la \rightarrow 0$ (right hand side picture).

Figure 2: The effective electron-positron interaction in the exchange channel;
the diagrams correspond to the generated and the instantaneous interactions.
The perturbative photon exchange with two different time orderings is also
depicted.

Table 1: The effective light front QED Hamiltonian matrix up to
second order in $e$ in the Fock space representation. The matrix elements
of the 'diagonal' (Fock state conserving) sectors are unrestricted in the
energy differences; the 'rest' (Fock state changing) sectors are squeezed
roughly in an energy band of width $\la$. Black dots correspond to zero
matrix elements in order $O(e^2)$. Instantaneous and disconnected
diagrams are not included.

\newpage
% MSR
\begin{figure}
$$
\fips{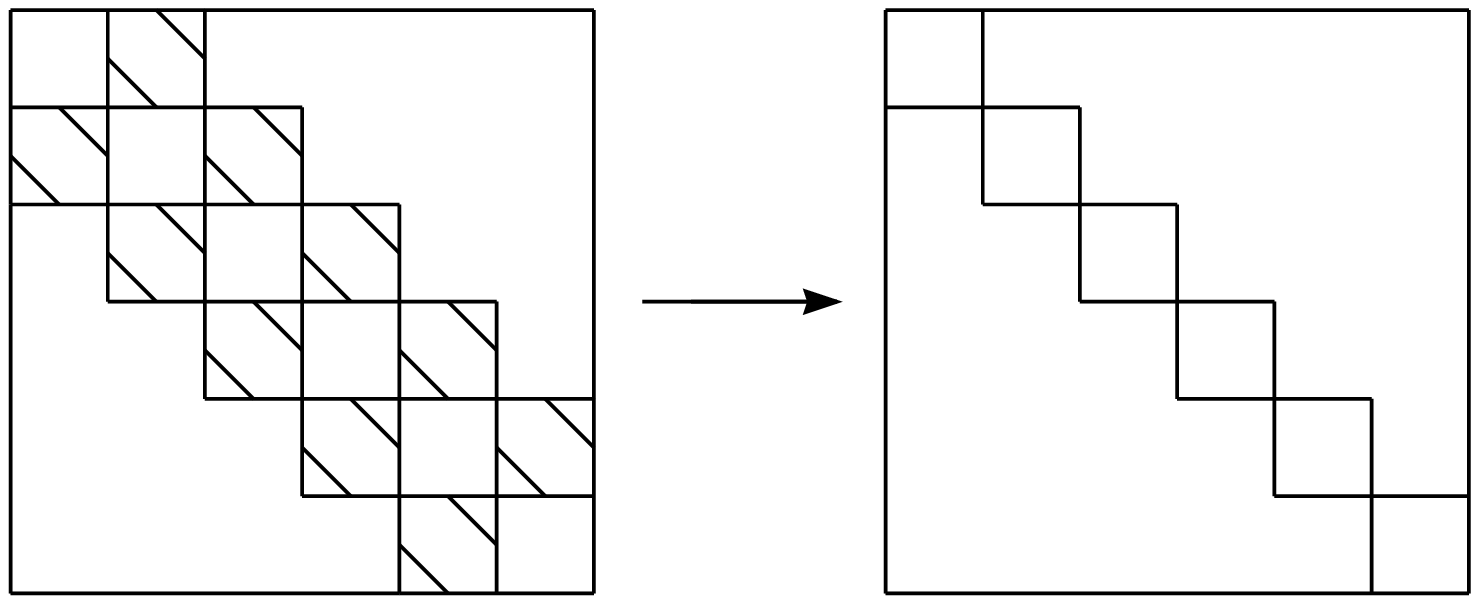}
\setlength{\unitlength}{0.240900pt}
\begin{picture}(0,0)
\put(-1500,780){\makebox(0,0){particle number}}
\put(-900,420){\makebox(0,0){$U(\La,\la)$}}
$$
\end{picture}
%\caption{Flow equations perform the block-diagonalization of 
%the bare Hamiltonian of the canonical theory $H_B(\La)$
%into blocks with the same number of quasiparticles. 
%For the finite value
%of $\la$ (after the unitary transformation
%$U(\La\rightarrow\infty,\la)$ is performed) the matrix elements 
%of 'particle number changing' sectors 
%are squeezed in the energy band $|E_i-E_j|<\la$ on the left hand side 
%picture and are eliminated completely as $\la\rightarrow 0$
%(that corresponds to $U(\La\rightarrow\infty,\la\rightarrow 0)$)
%on the right hand side picture. One ends up with the block-diagonal
%in 'particle number space' effective renormalized Hamiltonian.}
%\begin{center}
%Figure 4: Modified similarity renormalization of Hamiltonians
%\end{center}
\vspace{0.5cm}
$$
\begin{center}
Figure 1
\end{center}
\label{fig1}
\end{figure}

\newpage
% Table
\begin{table}
%\caption{The effective light front QED Hamiltonian matrix,
%renormalized to the second order, in the Fock space representation.
%The matrix elements of the 'diagonal' (Fock state conserving) sectors exist
%for any energy differences; the 'rest' (Fock state changing) sectors
%are sqeezed in the energy band 
%$\De_{p_ip_f}=|\sum p_i^--\sum p_f^-|<\la$; black dots correspond
%to the zero matrix elements to the order $O(e^2)$. Instantaneous diagramms
%are not included.}
\vspace{2cm}
\begin{tabular}{|r|c|c|c|c|c|} \hline
 & $|\ga>$ & $|e\bar{e}>$ & $|\ga\ga>$ & $|e\bar{e}\ga>$ 
& $|e\bar{e}e\bar{e}>$ \\ \hline
$|\ga>$ & \floadeps{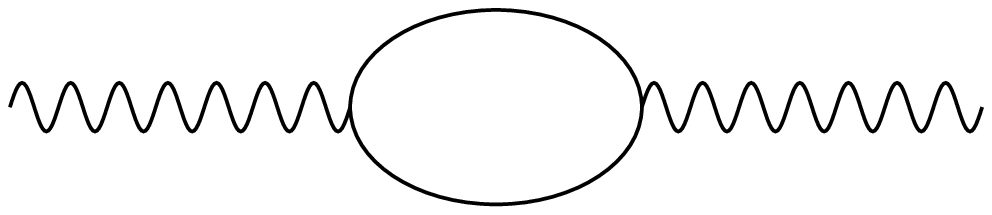} & \floadeps{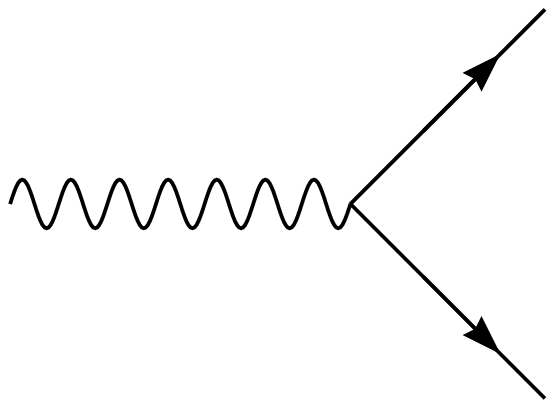} 
& \floadeps{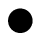} & \floadeps{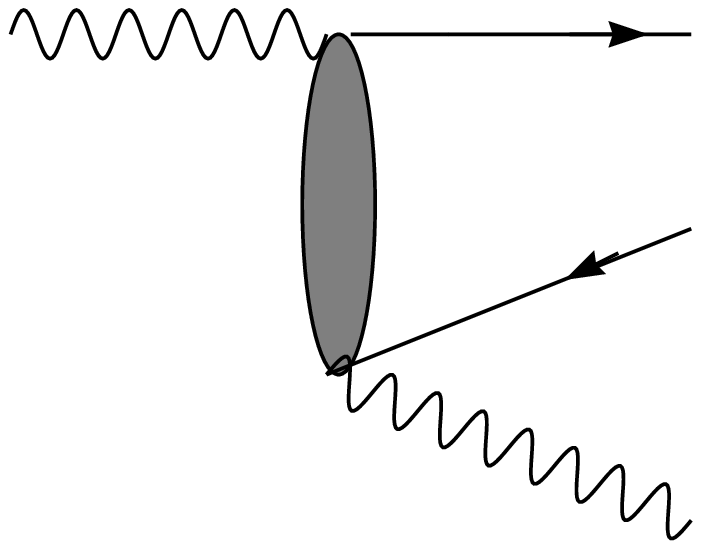} & \floadeps{table15} \\ \hline
$|e\bar{e}>$ & \floadeps{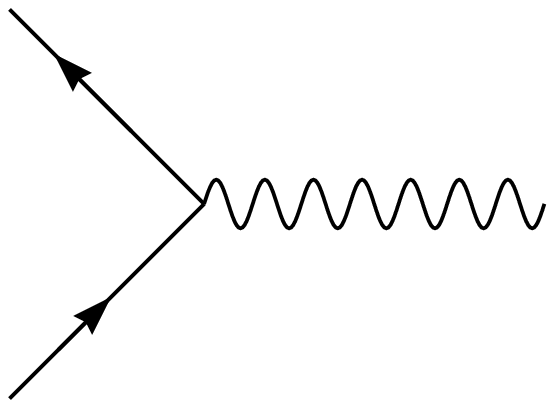} & \floadeps{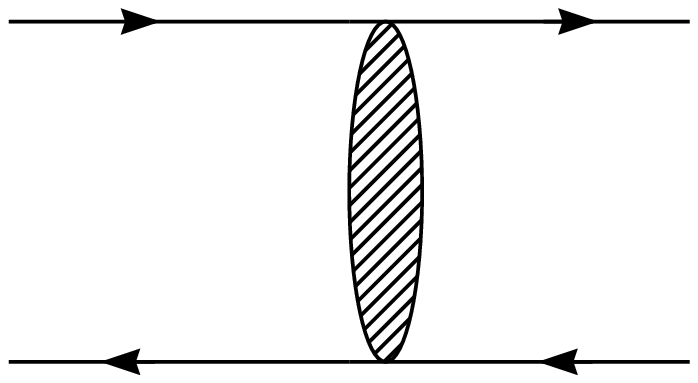} 
& \floadeps{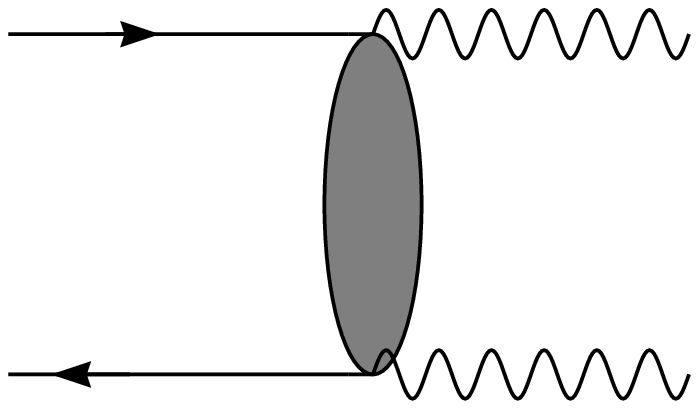} & \floadeps{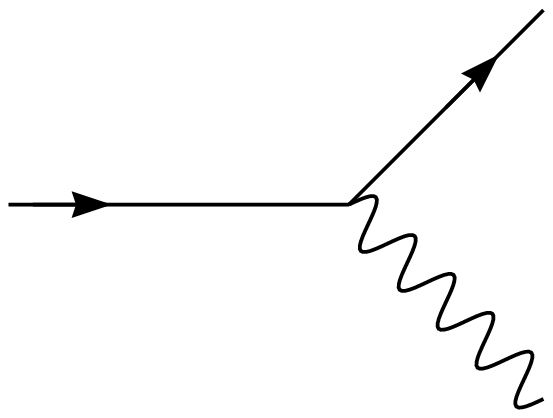} & \floadeps{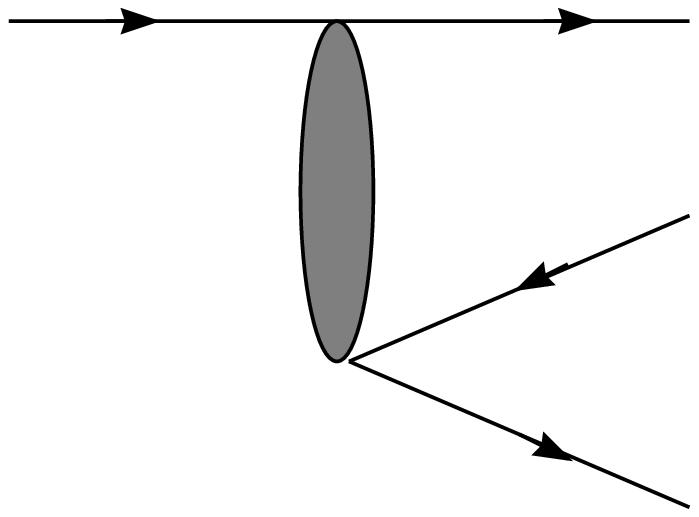} \\ \hline
$|\ga\ga>$ & \floadeps{table15} & \floadeps{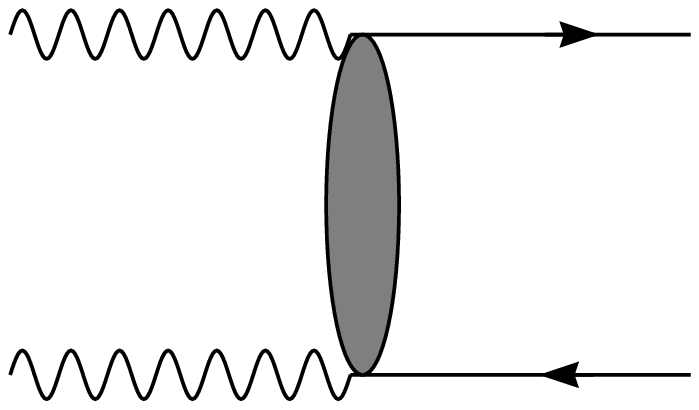} 
& \floadeps{table15} & \floadeps{tab12} & \floadeps{table15} \\ \hline
$|e\bar{e}\ga>$ & \floadeps{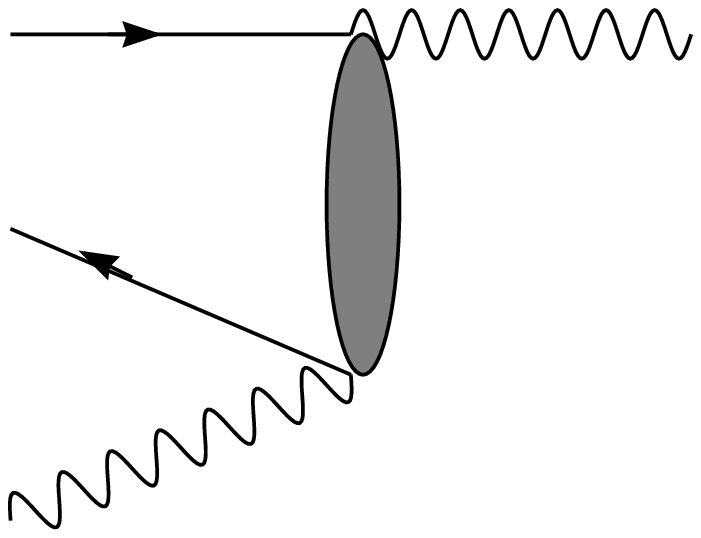} & \floadeps{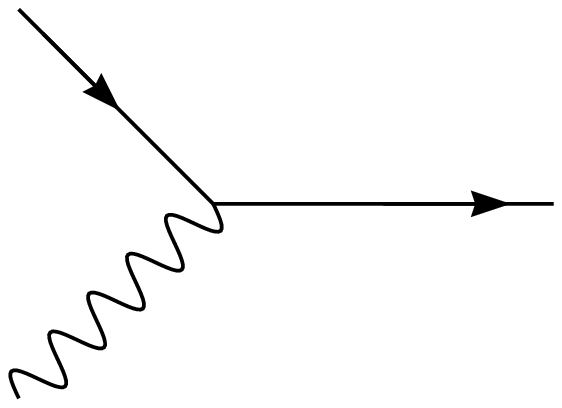} 
& \floadeps{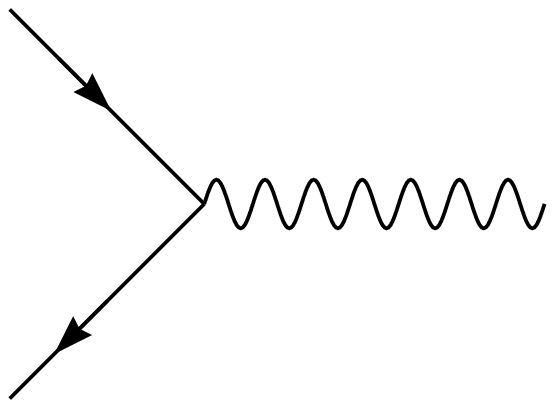} & \floadeps{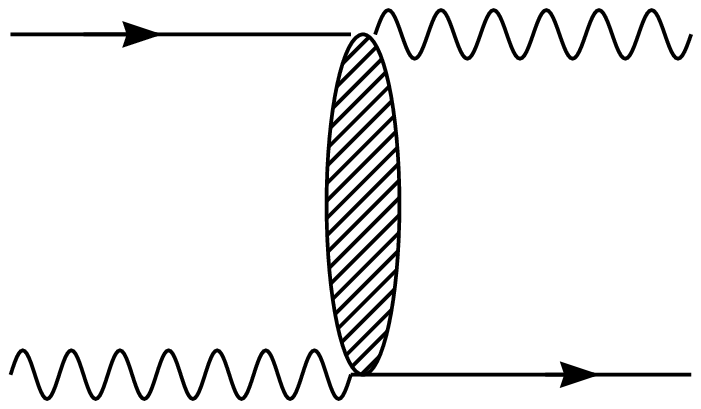} & \floadeps{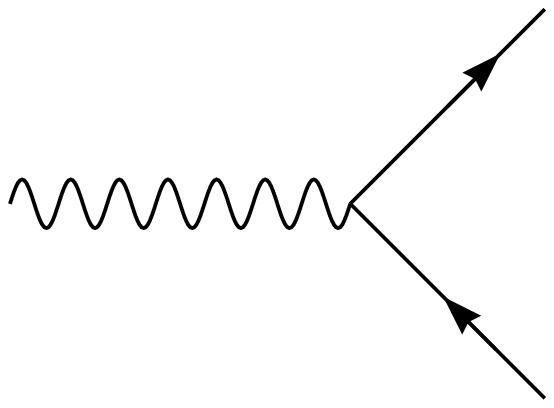} \\ \hline
$|e\bar{e}e\bar{e}>$ & \floadeps{table15} & \floadeps{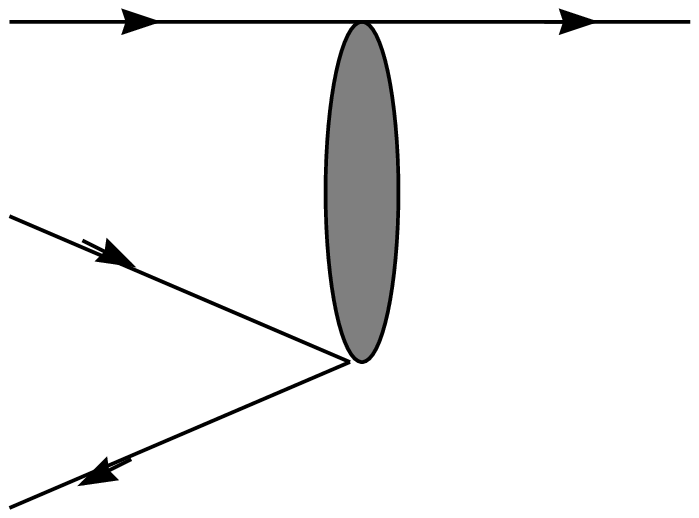} 
& \floadeps{table15} & \floadeps{tab43} & \floadeps{table22} \\ \hline
\end{tabular}
\vspace{1cm}
\begin{center}
Table 1
\end{center}
\label{table}
\end{table}

% Electron-positron interaction
\begin{figure}
\setlength{\unitlength}{1mm}
\begin{picture}(170,71)
\put(19,36){\makebox(56,34.61){ \loadeps{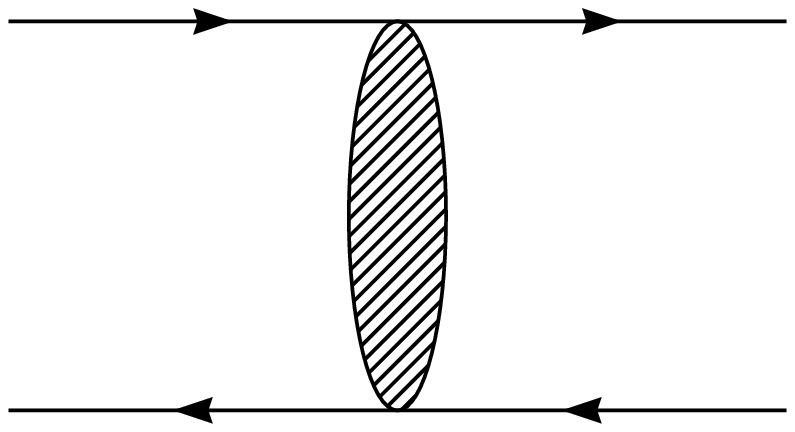} }}
\put(18,67){$p_1\;(x,k^\bot)$}
\put(50,67){$p_3\;(x',k'^\bot)$}
\put(18,38){$p_2\;(1\!-\!x,-k^\bot)$}
\put(50,38){$p_4\;(1\!-\!x',-k'^\bot)$}
  \put(75,36){\makebox(56,34.61){ \loadeps{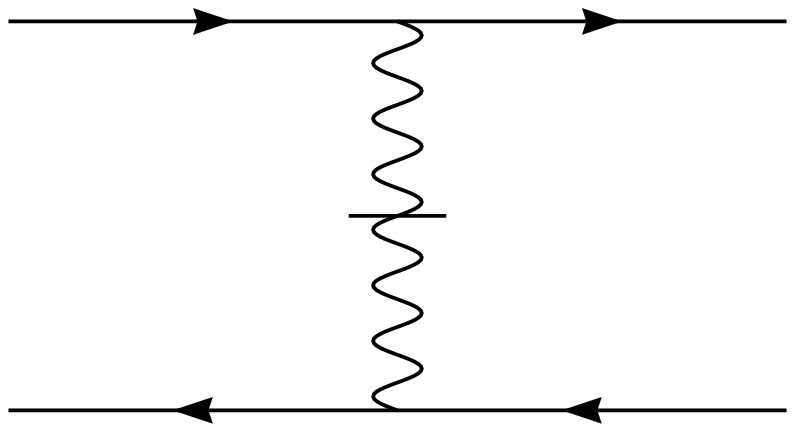} }}
\put(73,53.305){$+$}
\put(38,0){\makebox(56,34.61){ \loadeps{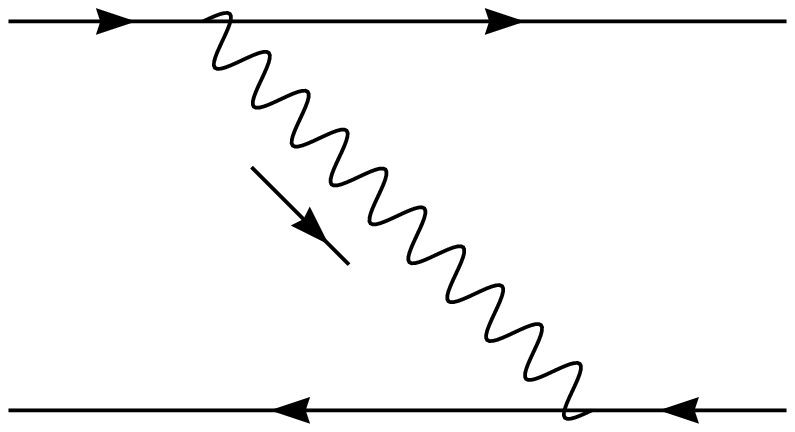} }}
  \put(94,0){\makebox(56,34.61){ \loadeps{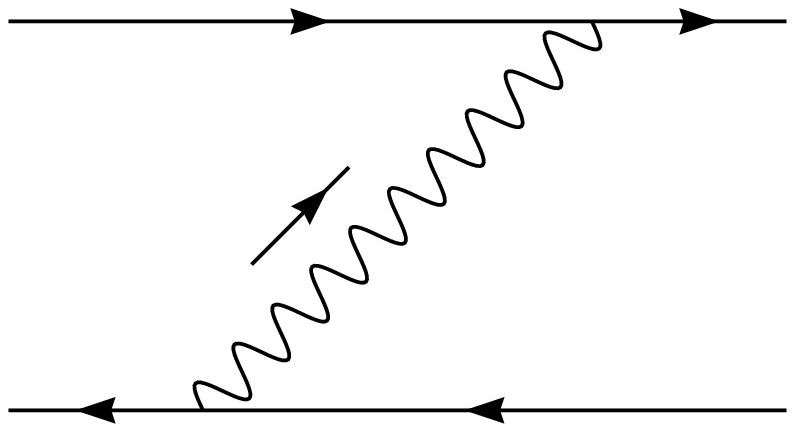} }}
\put(36,17.305){$+$}
\put(92,17.305){$+$}
\end{picture}
%\caption{The effective electron-positron interaction in the exchange channel; 
%the diagrams correspond to generated, instantaneous interactions and 
%two perturbative photon exchanges with respect to different time ordering.}
\vspace{0.5cm}
\begin{center}
Figure 2
\end{center}
\label{reneebarint}
\end{figure}

\end{document}